# Extended semi-Latin squares for use in field and glasshouse trials

E.R. WILLIAMS

Statistical Support Network, Australian National University, Canberra, Australia

**Abstract**
Semi-Latin squares have been extensively studied. They can be interpreted as a special case of latinized block designs where the number of columns ($s$) is equal to the number of replicates ($r$) in the design. Latinized row-column designs are frequently used in field and glasshouse trials when replicates are contiguous. These designs allow for the efficient adjustment of row and column effects within replicates. Here we define extended semi-Latin squares as the $r = s$ special case of latinized row-column designs and investigate optimality using the average efficiency factor.

**Author for correspondence:** Emlyn Williams, E-mail: emlyn@alphastat.net.

## 1. INTRODUCTION

Semi-Latin squares have been extensively studied for their combinatorial properties and are also used as experimental designs for field and glasshouse trials. For a detailed description of the construction, properties and application of semi-Latin squares, see Bailey (2017). We define a semi-Latin square design for $v = ks$ treatments as a $v \times s$ array of plots where the $v$ rows of the array can be grouped into $s$ subarrays of size $k \times s$ such that each subarray is a complete replicate of the treatments, i.e. the design is resolvable (John and Williams, 1995, chapter 4). Furthermore the $s$ long columns also constitute complete replicates of the treatments. In Figure 1 we present some examples of semi-Latin squares for $v = 8$ treatments with $k = 2$ and $s = 4$. The designs in Figures 1(a) and 1(b) were given by Bailey (2017) but we have transposed rows and columns and used numbers instead of letters. These semi-Latin squares exhibit different properties. In particular, if we consider the columns of size 2 within each replicate as incomplete blocks (16 of them), we can calculate the average efficiency factor ($E_{col}$) of each design. Here we define $E_{col}$ as the harmonic mean of the canonical efficiency factors of the incomplete block design (John and Williams, 1995, section 2.3). For the designs in Figure 1, the first has $E_{col} = 0$ since it is disconnected; the second has $E_{col} = 0.4636$. Bailey (2017) notes that semi-Latin squares can be considered as incomplete block designs and that it is important to choose a design that maximizes a theoretical efficiency criterion such as $E_{col}$. Hence the design in Figure 1(c) with $E_{col} = 0.5385$ would be preferred; this design is in fact optimal.

Williams (1986) introduced the class of latinized alpha designs based on the ideas of Harshbarger and Davis (1952) for latinized rectangular lattices. These were used for cotton variety trials where the long columns represent irrigation channels which usually show significant variation between channels. Hence it was important that treatments were spread evenly across the long columns. John and Williams (1995, section 4.6) generalized this latinization concept and defined the class of latinized block designs for $r$ contiguous replicates, each with $k$ rows and $s$ columns to include any type of incomplete block design with contiguous replicates. Williams *et al.* (2024, example 8.1) present an example of a forestry field trial where the trees in one of the long columns of the design were affected by water and could not be included for analysis. The latinization property ensured that no more than one plot of any of the provenances was affected. Semi-Latin squares can thus be viewed as latinized block designs with $r = s$ replicates. Using the Wilkinson and Rogers (1973) specification, the blocking structure for latinized block designs and hence the semi-Latin squares would be

columns + replicates/columns

```
1 7 2 8      1 3 5 7      4 8 3 5
5 3 6 4      2 4 6 8      2 1 7 6
6 4 5 3      3 6 1 2      8 4 5 3
2 8 1 7      4 8 7 5      6 7 1 2
3 5 4 6      5 2 8 1      3 2 6 8
7 1 8 2      6 7 4 3      1 5 4 7
8 2 7 1      7 1 2 4      5 3 2 1
4 6 3 5      8 5 3 6      7 6 8 4
   (a)          (b)          (c)
```
Figure 1. Semi-Latin squares for 8 treatments in 4 contiguous replicates of size 2 x 4 plots.

## 2. EXTENDED SEMI-LATIN SQUARES

When plots can be arranged in a rectangular array, latinized row-column designs (John and Williams, 1995, section 6.5) are commonly used in field and glasshouse trials. These designs are resolvable and also allow control of environmental variation in both the row and column direction within replicates. The Wilkinson and Rogers (1973) specification for the blocking structure would then be

columns + replicates/(rows+columns)

The construction of latinized row-column designs aims to maximize the average efficiency factor ($E_{rowcol}$) of the two-dimensional row-column design. It is also important to ensure that the component row and column average efficiency factors ($E_{row}$ and $E_{col}$) are adequate, as measured against theoretical upper bounds (John and Williams, 1995, chapters 2, 4 and 5). As mentioned in the previous section, semi-Latin squares are special cases of latinized block design when $r = s$. It is therefore of interest to investigate the extension of semi-Latin squares to incorporate rows within replicates as an extra blocking factor, i.e. a special case of latinized row-column designs.

As an example Bailey (1992, Figure 5) presents the semi-Latin square in Figure 2(a) (where again we have transposed rows and columns and used numbers rather than letters). This design has $E_{col} = 0.7$ which as Bailey observes, is optimal and hence achieves an upper bound defined as $U_{col}$. But without randomization of elements in columns within replicates, it becomes disconnected as a latinized row-column design with $E_{rowcol} = 0$. Bailey (1992, Figure 7) then gives the design in Figure 2(b) which has been randomized according to the blocking structure for a latinized block design. This now has $E_{rowcol} = 0.5034$ but can be improved on by designing in the row direction. CycDesigN generates the design in Figure 2(c), again with $E_{col} = 0.7$ but now has $E_{rowcol} = 0.5645$ compared with a (not necessarily tight) upper bound of $U_{rowcol} = 0.5688$.

The design in Figure 2(a) is an example of a special type of semi-Latin square called a Trojan square (Bailey, 2017) and will result in an optimal design, i.e. $E_{col} = U_{col}$. This design can be converted into an extended semi-Latin square by rotating the treatments in the last column of each replicate, as has been done in Figure 2(d). The canonical efficiency factors of this design are $2/3$ (multiplicity 4), $12/25$ ($\times 2$) and $41/75$ ($\times 8$) resulting in an average efficiency factor of $E_{rowcol} = 0.5645$, i.e. the same value as obtained from CycDesigN. Due to the general balance (Nelder, 1965) properties of Trojan squares, the first four canonical efficiency factors can be identified with columns within replicates, the next two with rows within replicates and the final eight with the interaction of rows and columns within replicates.

Because extended semi-Latin squares require *s* replicates, for most practical applications it means that *s* should not be too large. Also it is usually desirable to keep the overall dimensions of a field or glasshouse trial squarer in nature, meaning *k* is usually not greater than s. Thus we have used CycDesigN to construct extended semi-Latin squares for $k \leq s$ and up to $s = 7$. Average efficiency factors and upper bounds, as produced by CycDesigN, are summarized in Table 1. We note that in Tables 4 and 5 of Bailey (1992), the optimal but hypothetical Trojan square harmonic mean efficiency factors agree with $U_{col}$ in Table 1 for $s = 6, k = 2$ and $s = 6, k = 3$ respectively. CycDesigN allows the construction of optimal or near optimal row and column designs by either simultaneously performing row and column interchanges, or by first optimizing the design in one direction (i.e. the latinized block design) and then trying to maximize $E_{rowcol}$, holding the first-stage design fixed. In constructing Table 1 the two stage approach proved the best approach. In other words we first obtained an optimal semi-Latin square before carrying out interchanges within columns within replicates.

|  1 |  5 |  4 |  3 |  2 |   |  2 |  3 | 13 | 11 |  4 |   |  5 |  1 |  3 | 14 | 10 |   |  1 |  5 |  4 |  3 |  8 |
|---:|---:|---:|---:|---:|---|---:|---:|---:|---:|---:|---|---:|---:|---:|---:|---:|---|---:|---:|---:|---:|---:|
|  6 |  9 |  7 | 10 |  8 |   |  8 | 12 |  9 |  1 |  7 |   |  9 |  8 | 15 | 12 | 13 |   |  6 |  9 |  7 | 10 | 14 |
| 11 | 13 | 15 | 12 | 14 |   | 14 | 10 |  5 |  6 | 15 |   |  7 |  2 |  4 | 11 |  6 |   | 11 | 13 | 15 | 12 |  2 |
|  2 |  1 |  5 |  4 |  3 |   | 12 | 15 |  7 |  4 | 13 |   |  6 |  5 | 11 |  2 |  8 |   |  2 |  1 |  5 |  4 |  9 |
|  7 | 10 |  8 |  6 |  9 |   |  6 |  1 |  3 |  9 |  2 |   |  3 | 10 |  1 |  9 |  4 |   |  7 | 10 |  8 |  6 | 15 |
| 12 | 14 | 11 | 13 | 15 |   |  5 |  8 | 11 | 14 | 10 |   | 12 | 15 |  7 | 13 | 14 |   | 12 | 14 | 11 | 13 |  3 |
|  3 |  2 |  1 |  5 |  4 |   | 13 |  9 |  4 | 15 |  3 |   |  1 | 14 |  2 |  3 |  9 |   |  3 |  2 |  1 |  5 | 10 |
|  8 |  6 |  9 |  7 | 10 |   |  7 |  2 |  8 | 10 |  6 |   |  4 |  6 | 12 |  5 | 11 |   |  8 |  6 |  9 |  7 | 11 |
| 13 | 15 | 12 | 14 | 11 |   |  1 | 11 | 12 |  5 | 14 |   | 13 |  7 | 10 |  8 | 15 |   | 13 | 15 | 12 | 14 |  4 |
|  4 |  3 |  2 |  1 |  5 |   |  3 |  6 | 14 |  2 |  8 |   | 14 | 11 |  9 | 10 |  1 |   |  4 |  3 |  2 |  1 |  6 |
|  9 |  7 | 10 |  8 |  6 |   |  9 |  4 | 10 | 12 | 11 |   |  2 | 13 |  6 |  4 |  5 |   |  9 |  7 | 10 |  8 | 12 |
| 14 | 11 | 13 | 15 | 12 |   | 15 | 13 |  1 |  7 |  5 |   | 15 |  3 |  8 |  7 | 12 |   | 14 | 11 | 13 | 15 |  5 |
|  5 |  4 |  3 |  2 |  1 |   | 11 |  7 |  2 |  3 | 12 |   |  8 | 12 | 13 |  1 |  7 |   |  5 |  4 |  3 |  2 |  7 |
| 10 |  8 |  6 |  9 |  7 |   | 10 | 14 | 15 |  8 |  1 |   | 10 |  9 | 14 |  6 |  3 |   | 10 |  8 |  6 |  9 | 13 |
| 15 | 12 | 14 | 11 | 13 |   |  4 |  5 |  6 | 13 |  9 |   | 11 |  4 |  5 | 15 |  2 |   | 15 | 12 | 14 | 11 |  1 |
| (a) | | | | |   | (b) | | | | |   | (c) | | | | |   | (d) | | | | |

Figure 2. Semi-Latin squares, (a) and (b) and extended semi-Latin squares, (c) and (d) for 15 treatments in 5 contiguous replicates of size $3 \times 5$ plots.

| s | k | $E_{col}$ | $U_{col}$ | $E_{rowcol}$ | $U_{rowcol}$ |
|---|---|---|---|---|---|
| 3 | 2 | 0.5556 | 0.5556 | 0.310078 | 0.381239 |
| 3 | 3 | 0.615385 | 0.64 | 0.389355 | 0.468361 |
| 4 | 2 | 0.538462 | 0.538462 | 0.388889 | 0.421675 |
| 4 | 3 | 0.709677 | 0.709677 | 0.53457 | 0.537114 |
| 4 | 4 | 0.75 | 0.757576 | 0.576923 | 0.584416 |
| 5 | 2 | 0.529412 | 0.529412 | 0.427006 | 0.441533 |
| 5 | 3 | 0.7 | 0.7 | 0.564498 | 0.568839 |
| 5 | 4 | 0.780822 | 0.780822 | 0.619706 | 0.627006 |
| 5 | 5 | 0.810413 | 0.81203 | 0.650955 | 0.657534 |
| 6 | 2 | 0.513333 | 0.52381 | 0.442869 | 0.450891 |
| 6 | 3 | 0.692155 | 0.693878 | 0.579481 | 0.585568 |
| 6 | 4 | 0.767104 | 0.775281 | 0.643221 | 0.650917 |
| 6 | 5 | 0.814233 | 0.822695 | 0.67920 | 0.686859 |
| 6 | 6 | 0.844221 | 0.844828 | 0.704421 | 0.708502 |
| 7 | 2 | 0.52 | 0.52 | 0.458019 | 0.460243 |
| 7 | 3 | 0.689655 | 0.689655 | 0.593826 | 0.598007 |
| 7 | 4 | 0.771429 | 0.771429 | 0.659581 | 0.665848 |
| 7 | 5 | 0.819277 | 0.819277 | 0.698402 | 0.705191 |
| 7 | 6 | 0.850622 | 0.850622 | 0.723991 | 0.729865 |
| 7 | 7 | 0.866471 | 0.867209 | 0.739553 | 0.746114 |

Table 1. Average efficiency factors and upper bounds for the column component and overall row-column design for extended semi-Latin squares.

## 3. UPPER BOUNDS

For some combinations of *r*, *k* and *s*, theory provides optimal designs but for the majority of cases required in practice, computer packages are used to obtain designs with optimal or near-optimal average efficiency factors. Then it is important to have tight upper bounds for these quantities; for example if an iterative computer optimization process can get within one or two percent of an upper bound, then the design can often be accepted for use in practice. It is worth pointing out that the theoretical design efficiency factor is a contributor to the overall efficiency of an incomplete block design, relative to say a randomized complete block design.

John and Williams (1995, section 4.10) give upper bounds for resolvable block designs in general. But for $k \geq s$, the latinization requirement places a restriction on $U_{col}$ (Williams and John, 1993) resulting in the tighter bounds for $k = s$ in Table 1. In particular for $k = s$ this bound is given by

$$U_{col} = (s+1)^2(s-2)/(s^3 + s^2 - 3s - 2) \qquad (1)$$

In the Appendix, however, we have derived a new quantity, namely

$$W_{col} = s(s+1)(s-2)/(s^3 - 3s + 1) \qquad (2)$$

which we show is an upper bound for $s = 6$.

In Table 2, (2) is seen to be tighter than (1) for the Table 1 designs. We note that for $s = 6$, the latinized block design achieves $W_{col}$ and hence is optimal. For the other values of *s*, the best designs are still below $W_{col}$. But for these smaller designs we would expect the computer search to locate the optimal designs so it seems that (2) is still not tight for $s \neq 6$.

| $k = s$ | $E_{col}$ | $W_{col}$ | $U_{col}$ |
|---|---|---|---|
| 3 | 0.615385 | 0.631579 | 0.640000 |
| 4 | 0.75 | 0.754717 | 0.757576 |
| 5 | 0.810413 | 0.810811 | 0.812030 |
| 6 | 0.844221 | 0.844221 | 0.844828 |
| 7 | 0.866471 | 0.866873 | 0.867209 |

Table 2. Average efficiency factors and bounds for the $k = s$ designs considered in Table 1.

## 4. DISCUSSION

The construction of optimal designs for use in practice has greatly benefitted from background theory from mathematical areas such as combinatorics and group theory. But such approaches can only address quite a small number of combinations of treatments, block sizes and replicates in the ranges likely to be of most use for field and glasshouse trials. Use of computer search algorithms has helped to fill in the gaps; this was the case when the class of alpha designs was developed (Patterson and Williams, 1976). These days, with very much more computer power available, computer search techniques are available for the generation of a wide range of optimal or near-optimal designs and including many different design features such as incomplete blocking in one or two directions, latinized designs and spatial designs just to name some. With computer search techniques, the availability of tight upper bounds is very important so that users can know how close they are to theoretically derived optima. In practice it is often sufficient to use a near-optimal design; this is especially the case for designs in plant and tree breeding where there can be many hundreds of treatments. For the small designs considered in Table 1, however, CycDesigN will find the optimal design quite quickly and so the (often not tight) upper bounds are generally less important.

In this paper we have focussed on parameter combinations with $k \leq s$ as this is likely to result in squarer dimensions for any experimental layout; in other words the Figure 3 structure is usually to be preferred over that in Figure 2. In constructing the designs we have used a two-stage approach where firstly we obtain an optimal or near-optimal latinized block design and then further develop this into an efficient latinized row-column design. For the range of values of $s$ and $k$ under study here, we have found that the best row-column design has always been developed from a semi-Latin square. Although there was one situation for $s = k = 4$ where the best latinized row-column design was also achieved when the component latinized block design was not a semi-Latin square.

The design for $s = k = 6$ in Table 2 is of particular interest. This design is well known (see Bailey *et al*., 2020) and it is the only case we found where the constructed design achieves $W_{col}$. In fact Soicher (2013) calls this design 'remarkable' and we agree, especially since six is the only number where there are no mutually orthogonal Latin squares.

# APPENDIX
## A proposed upper bound for latinized block designs for $v = s^2$ treatments in blocks of size *s* and *s* replicates.

In this situation there are also $b = s^2$ blocks. The $b \times b$ information matrix for the dual design is given by
$$A_d = sI_{s^2} - N'N$$
where $N$ is the $v \times b$ incidence matrix and $I_{s^2}$ is the identity matrix of size $s^2$ (John and Williams, 1995, chapter 2)

As discussed by Williams and John (1993), the concurrence matrix, $N'N$ for the dual design can be partitioned into $s \times s$ submatrices with each diagonal submatrix equal to $sI_s$ and the off-diagonal submatrices having row and column sums equal to $k$. Latinized block designs then have the further restriction that the diagonal elements of these off-diagonal submatrices are equal to zero. Again following Williams and John (1993), we focus on the class of latinized block designs with the off-diagonal elements of the off-diagonal submatrices of $N'N$ differing by not more than one. For our situation, this means that these submatrices will have one off-diagonal element equal to 2 in each row and column, with the rest equal to 1. Then $A_d^* = \frac{1}{s}A_d$ can be written as

$$A_d^* = \frac{s^2-s-1}{s^2}I_{s^2} - P_0 + \frac{1}{s}P_1 + \frac{1}{s}P_2 - \frac{1}{s^2}Q$$

where $P_0 = \frac{1}{s^2}I_{s^2}$, $P_1 = \frac{1}{s}I_s \otimes J_s$, $P_2 = \frac{1}{s}J_s \otimes I_s$ are projection matrices (James and Williams, 2024) and $J_s$ is a matrix of ones. Q is a matrix that can be partitioned into submatrices where the diagonal submatrices are zero and the off-diagonal submatrices have zero elements except for a single off-diagonal value of 1 in each row and column.. The matrix $A_d^*$ is singular but it can be shown by multiplying out that the inverse of $A_d^* + P_0$ is given by

$$A_d^{**} = \{s^3(s-1)I_{s^2} - s^2(s-1)^2 P_0 - s^2(s-1)P_1 - s^2(s-1)P_2 + s^2 Q^*\}/\{s^3(s-2)\}$$
where
$$Q^* = \{(s^2-s-1)I_{s^2} - Q\}^{-1}\{s(s-1)Q - sI_{s^2} - s^2 P_0 + sP_1 + sP_2\}$$
Projection matrix properties help here, e.g.
$P_1 P_2 = P_0$, $P_0 P_i = P_0$ and $QP_i = sP_0 - P_i$ for $i = 0,1,2$.

The average efficiency factor of the latinized block design can then be written as
$$E_{col} = (s^2 - 1)/trace(A_d^{**} - P_0)$$
i.e. the harmonic mean of the canonical efficiency factors.
We can evaluate part of $trace(Q^*)$ namely,
$$trace(Q^*) = 1 - trace(\{(s^2 - s - 1)I_{s^2} - Q\}^{-1}\{sI_{s^2} - s(s-1)Q\})$$
For the designs whose properties are listed in Table 1, $trace(Q^*) \geq 0$. Hence dropping the $Q^*$ term from $A_d^{**}$ will yield an upper bound for $E_{col}$. After simplification, this reduces to the expression (2) given for $W_{col}$.
It is interesting that when $s = 6$, $Q^* = Q$ for Soicher's (2013) 'remarkable' design, i.e. $trace(Q^*) = 0$.